\documentclass[conference]{IEEEtran}
\usepackage{graphicx}
\usepackage{epstopdf}
\usepackage{mathrsfs}
\usepackage{mathrsfs}
\usepackage{mathrsfs}
\usepackage{amssymb}
\usepackage{amsthm}
\usepackage{amsmath}
\usepackage{amsfonts}
\usepackage{mathrsfs}
\usepackage{amssymb}
\usepackage{subfigure}
\usepackage{bm}
\usepackage{caption}
\usepackage{stfloats}
\usepackage{epsfig}
\usepackage{algorithm}
\usepackage{algorithmic}
\usepackage[dvips]{color}
\usepackage{subeqnarray}
\usepackage{cases}
\usepackage{cite}

\begin{document}

\title{Performance Analysis for the CMSA/CA Protocol in UAV-based IoT network}

\author{\IEEEauthorblockN{Xianzhen Guo, Bin Li, Kebang Liu}\\
\IEEEauthorblockA{Department of Communication Engineering, Northwestern Polytechnical University, Xi'an, China 710072}
}

\maketitle

\begin{abstract}

UAV-based base station (UBS) has played an important role in the air-ground integration network due to its high flexibility and nice air-ground wireless channels. Especially in Internet of Things (IoT) services, UBS can provide an efficient way for data collection from the IoT devices.  
However, due to the continuous mobility of UBS, the communication durations of devices in different locations with the UBS are not only time-limited, but also vary from each other.  
Therefore, it is a challenging task to analyze the throughput performance of the UAV-based IoT network. Accordingly, in this paper, we consider an air-ground network in which UAV flies straightly to collect information from the IoT devices based on CSMA/CA protocol. An analytical model analyzing the performance of this protocol in the network is proposed. In detail, we set up the system model for the network, and propose a new concept called quitting probability. Then, a modified Markov chain model integrating the quitting probability is introduced to describe the transmission state transition process and an accurately theoretical analysis of saturation throughput is given. In addition, the effects of the network parameters are discussed in the simulation section.
\end{abstract}

\begin{IEEEkeywords}
Air-Ground integration network, CSMA/CA, IoT, Markov chain model, UAV
\end{IEEEkeywords}

\IEEEpeerreviewmaketitle

\begin{figure}[b]
\vspace{-4mm}
\footnotesize{This work is partially supported by National Natural Science Foundation of China (Nos. 61601365, 61571370), Natural Science Basic Research Plan in Shaanxi Province (No. 2019JM-345), National Civil Aircraft Major Project of China (No. MIZ-2015-F-009), China Postdoctoral Science Foundation(No. 2018M641020)
}
\end{figure}

\IEEEpeerreviewmaketitle

\section{Introduction}
Wireless communication with unmanned aerial vehicles (UAVs) has been adopted widely in air-ground integration network for military, public and emergency applications \cite{b1}. 
Compared with the terrestrial communication system, UAV-based air-ground network is not only easy to deploy, but also has better wireless channels \cite{b2}. 
Therefore, numerous UAV-based wireless communication systems have been developed \cite{d1,d2,d3,d4}. 
One typical application is the UAV-based data collection network in the air-ground IoT system, where UAV is deployed as a UBS flying over the IoT devices to collect the information. 
In this way, the system avoids the complex routing design and improves the efficiency of data collection greatly. 
However, there are still some challenges in developing UAV-based data collection system.

One of the challenges is that it is difficult to analyze the performance of the MAC protocol performance for the mobility of the UBS. 
Generally, the fixed-wing UAV is usually a priority in data collection scenario for its high altitude and long endurance. 
However, the UAV of this type cannot hover over a predetermined location and must fly continuously, causing the time-varying channels between the UAV and the devices. 
Moreover, the UAV is usually equipped with directional antennas for energy-efficient transmission, making a circular signal coverage on the ground. In this case, the devices in different locations can communicate with UBS for different durations.
These characteristics make it difficult to develop an accurate and efficient theoretical model to analyze the performance of a UAV-based communication system. 
Given the wide application of the CSMA/CA in data collection network, we focus on the performance analysis of CSMA/CA considering the continuous mobility of UBS. 

Some work has been made to analyze the performance of the UAV-based data collection system. 
\cite{ba} analyzes the influence of the distance between two UAVs on the system performance, but the UAV's mobility was not taken into consideration. 
In \cite{b3}, the location of UAV is updated periodically to obtain the optimized locations of UAV. However, it only involves the rotor crafts which could hover over the predetermined locations to collect the information. 
\cite{b6} gives a simple but extremely accurate model to calculate the throughput of CSMA/CA protocol using Markov chain model, and \cite{b7} extends the multi-dimensional Markov chain model by considering the impacts of both non-ideal channel and capture effects. 
Unfortunately, both of them can only be applied in terrestrial communication system with fixed base station.

In this paper, we propose a new quitting probability based on the CMSA/CA protocol to deal with the communication heterogeneity among the devices caused by UBS's mobility. 
In detail, we divide the devices into different clusters according to their communication durations with the UBS. 
Then, a modified Markov chain model integrating the quitting probability is introduced to describe the transition of the devices' state and an accurate analytical model of saturation throughput is given.

Our main novelty and contributions are given as follows.
\begin{itemize}
\item We analyze the characters of the UAV-based IoT network and propose a new quitting probability in the Markov chain model to model the devices access process where the different locations among the devices are considered.
\item We set up an accurate theoretical analysis model and calculate the saturation throughput of CSMA/CA protocol in UAV-based IoT network.
\end{itemize}

The rest of this paper is organized as follows. System model is presented in Section~\ref{section:2}. Section~\ref{section:3} analyzes the Markov model process and gives the computation of throughput. Numerical results are shown in Section~\ref{section:5}. Finally, Section~\ref{section:6} concludes the paper.

\section{System Model}\label{section:2}

\subsection{UAV-based Air-ground Network Model}
\begin{figure}[t]
	\centering
	\includegraphics [width=7cm]{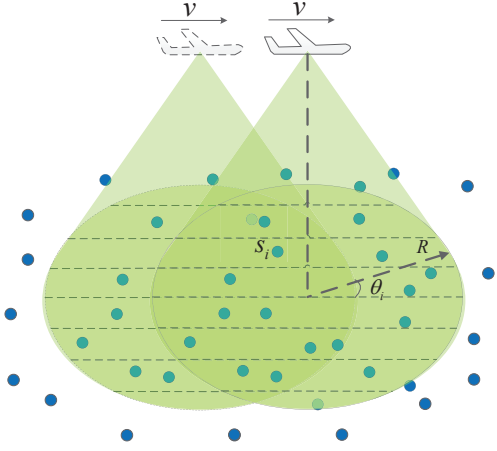}
	\caption{System model.}
	\label{system model}
\vspace{-5mm}
\end{figure}

We consider a scenario where UBS collects the information from a set of IoT devices distributed normally on the 2-D plane. Let $({x_u},{y_u},{z_u})$ and $({x_i},{y_i},{z_i})$ denote the locations of the UBS and $s_i,  i=1,2,...,K$, where $K$ is the total number of devices in the UAV's coverage. The UBS is assumed to move in a straight trajectory over the devices which try to access the UBS for data uploading. The access process between the devices and the UBS follows CSMA/CA protocol.

During the data collection, for energy efficiency, the UBS deploys the directional antennas for communications and forms a circular coverage on the ground, as shown in Fig.~\ref{system model}. Compared with the traditional communication system with terrestrial infrastructures, the communication durations between UBS and devices are time limited because of the UBS's mobility. 
The devices turn into the active mode and ready to send the information to the UBS once they are inside the coverage while the wireless links will break when the UBS flies away. In addition, the UBS-Devices communication durations are also different under the straight trajectory because of the different locations of the devices. The communication duration for device $s_i$ can be expressed as:
\begin{equation}
{T_i} = \frac{{2R\cos {\theta _i}}}{v},
\end{equation}
where $R$ is the radius of the UAV coverage, ${\theta _i} \in (0,\pi /2)$ is the locations of the device and can be calculated as ${\theta _i} = \arcsin ({x_i}/R)$ and $v$ is the velocity of UAV. How to generate the cluster will be described in detail in Section~\ref{section:2}-D.

\subsection{Markov Chain Model}
To analyze the network performance, we propose a Markov chain model to explore the access process of the devices in the network. We divide all the devices in the coverage into $N$ band-shaped clusters along the flight trajectory. In each cluster $\mathbb{C}_i$ ($i=1,...,N$), there are $n_i$ devices and each of them is assumed to always have a packet available for transmission. Assume that all the packet collide with constant and independent probability regardless of the number of retransmission suffered. 

Generally, the CSMA/CA access scheme employs a binary exponential backoff technique. During the transmission, the backoff time is uniformly chosen in the range of $(0,{W_{i,j}})$, where ${W_{i,j}}$ is the contention window in the $j$th backoff stage for the devices in cluster $\mathbb{C}_i$. For the first access round, the $W_{i,0}$ equals to $C{W_{\min }}$. After each failed access, contention window will be doubled, up to the maximum $C{W_{\max }}$. Therefore, the size of contention window in stage $j$ is ${W_{i,j}}{\rm{ = }}{2^j}{W_{i,0}},~0 \le j \le {L}$, where $L$ is the retry limit.

\begin{figure}[ht]
	\centering
	\includegraphics [width=8.5cm]{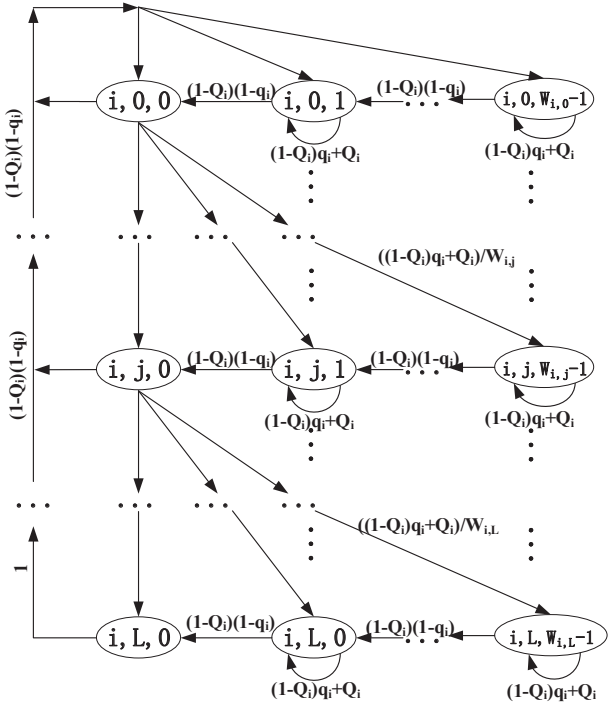}
	\caption{Markov chain model.}
	\label{Markovchainmodel}
\end{figure}

To clearly describe Markov chain model, we introduce three coefficients ($i$, $j$ and $b(t)$) to represent the Markov process. Here $b(t)$ represents the backoff counter for a certain device at time $t$. The backoff counter decreases if the channel is sensed idle and stops when the channel is busy. Let $s(t)$ be the stochastic process representing the devices' backoff stage $(j=0,...,m,s(t)=j)$ at time $t$. The backoff stage changes after any unsuccessful access transmission until to the maximal value. A three-dimensional Markov process with $i$, $j$ and $b(t)$ can be defined and the detailed stage transition process is shown in Fig.~\ref{Markovchainmodel}, where the ellipses with parameters ($i$, $j$ and $b(t)$) represent the states of this Markov process. The arrows show the direction of the state transmission. The formulas along the arrows, like $(1-Q_i)(1-q_i), (1-Q_i)q_i+Q_i$, are the probabilities for one-step transmission. $Q_i$ is the quitting probability and the details are given in Section \ref{quitting}. $q_i$ is the probability that the devices in the $i$th cluster sense the channel busy in a backoff stage and it is derived in Section \ref{Markov}.

According to Fig.~\ref{Markovchainmodel}, the one-step transmission probabilities of the Markov process is shown below.

\begin{subequations}\label{eq1_1}
	\begin{align}
		&  \{\begin{array}{l}
P_{i,j,k|i,j,k + 1} =(1-Q_i)(1-q_i)\\
j \in [0,L],~k \in [0,W_{i,j} - 2]
\end{array}, \\
		& \{\begin{array}{l}
{P_{i,j,k|i,j,k}} = (1 - {Q_i}){q_i} + {Q_i}\\
~j \in [0,{L}],~k \in [1,{W_{i,j}} - 1]
\end{array},\\
		& \{\begin{array}{l}
{P_{i,0,k|i,j,0}} = \frac{{(1 - {Q_i})(1 - {q_i})}}{{{W_{i,0}}}}\\
~j \in [0,{L} - 1],~k \in [1,{W_{i,0}} - 1]
\end{array},\\
		& \{\begin{array}{l}
{P_{i,0,k|i,{L},0}} = \frac{1}{{{W_{i,0}}}}\\
~k \in [1,{W_{i,0}} - 1]
\end{array},\\
		& \{\begin{array}{l}
{P_{i,j,k|i,j - 1,0}} = \frac{{(1 - {Q_i}){q_i} + {Q_i}}}{{{W_{i,j}}}}\\
~j \in [1,{L}],~k \in [0,{W_{i,j}} - 1],
\end{array}
	\end{align}
\end{subequations}
where $P_{i_1,j_1,k_1|i_0,j_0,k_0}$ is a short notation and it is defined as
\begin{small}
\begin{equation}
\begin{split}
&P_{i,j_1,k_1|i,j_0,k_0}~=\\&~P\{i,~s(t+1)=j_1,~b(t+1)=k_1~|~i,~s(t)=j_0,~b(t)=k_0\}.
\end{split}
\end{equation}
\end{small}

The equation (\ref{eq1_1}a) means that the backoff counter deceases in the beginning of each time slot with the probability that the device stays in the coverage area and the channel is idle at the same time. (\ref{eq1_1}b) tells us the probability that the device stays in the same stages because it quits the coverage or the channel is sensed busy. (\ref{eq1_1}c) and (\ref{eq1_1}d) represent the probability that the backoff stages return to 0 from the former stages. For devices in stages $(i,j,0)$, they will transmit to $(i,0,k)$ after a successful transmission or quit the network. For the stages in $(i,L,0)$, they will return to stages 0 anyway. The last equation gives the probability of rescheduling a contention state after an unsuccessful transmission.

\subsection{Quitting Probability}\label{quitting}

Since the mobility of the UBS, the devices on the ground will quit the access process if it is removed from the signal coverage. To model this special scenario, a new concept called quitting probability is proposed to define this quitting behavior. Since the devices have different communication durations with the UBS, their packets for transmission will go through the Markov process for different times. The longer they stay in UBS's coverage, the higher probability they can transmit packets successfully because they have higher opportunities to retry the network access.

To calculate the quitting probability of the devices in cluster $\mathbb{C}_i$, we first consider a situation where a packet traverses all the backoff stages, which indicates that the device does not quit the network in the whole access process. In this case, the transmission state transfers to state $(i,L,0)$. Here we use $P_b$ to denote the stationary probability of state $(i,L,0)$. Obviously, $(1-P_b)$ represents the probability that the packet quits the access process during the process.

While for cluster $\mathbb{C}_i$, the quitting probability ${Q_i}$ can be defined as the probability that the device quits the access process after $m_i$ times, where $m_i$ shows the maximal access times for cluster $i$. Accordingly, ${Q_i}$ can be shown as
\begin{equation}
{Q_i} = {(1 - {P_b})^{{m_i}}},
\end{equation}
where $m_i = \frac{{{T_i}}}{\Delta }$. ${\Delta}$ is the time duration after the packet traverses all the stages and the details can be find in Section \ref{cluster}.

By introducing the quitting probability into the Markov chain model, we can insight the transmission process in the UAV-based network.

\subsection{The Division of the Devices Clusters}\label{cluster}
\newcounter{tmepEq}
\setcounter{tmepEq}{\value{equation}}
\setcounter{equation}{18}
\begin{figure*}[!b]
\hrulefill
\begin{equation}\label{long_1}
{q_i} = 1 - \prod\limits_{h = 1}^{i - 1} {(\sum\limits_{n = 0}^\infty  {{f_h}(n)} {{(1 - {\tau _h})}^{{n_h}}})(\sum\limits_{n = 1}^\infty  {{f_i}(n)} {{(1 - {\tau _i})}^{n - 1}}} )\prod\limits_{h = i + 1}^N {(\sum\limits_{n = 0}^\infty  {{f_h}(n)} {{(1 - {\tau _h})}^{{n_h}}})}
\end{equation}\vspace{-0mm}
\begin{equation}\label{long_2}
{b_{i,j,k}} = \frac{{{W_{i,j}} - k}}{{{W_{i,j}}}}\left\{ {\begin{array}{*{20}{l}}
	{(1 - {q_i})(1 - {Q_i})\sum\limits_{j = 0}^{{L_j} - 1} {{b_{i,j,0}} + {b_{i,{L},0}},~j = 0} }\\
	{[{q_i}(1 - {Q_i}){\rm{ + }}{Q_i}]{b_{i,j - 1,0}},~j \in [1,L]}
	\end{array}} \right.
\end{equation}
\end{figure*}
\setcounter{equation}{\value{tmepEq}}

\setcounter{equation}{21}
\begin{figure*}[!b]
\begin{small}
	\begin{equation}\label{long_3}
	{b_{i,0,0}} = \frac{{2{P_a}{P_b}}}{{2{W_{i,0}}{P_{eq}}[{P_a}(1 - {{(2{P_{eq}})}^{{L}}}) - (1 - P_{eq}^{{L}}){P_b}] + 2{P_b}(1 - P_{eq}^{{L} + 1}) + {P_b}[(1 - {q_i})(1 - {Q_i})(1 - P_{eq}^{{L}}) + {P_a}P_{eq}^{{L}}]({W_{i,0}} - 1)}}
	\end{equation}
\end{small}
\end{figure*}
\setcounter{equation}{\value{tmepEq}}

As mentioned before, the devices in different locations have different time to access the UBS. To insight the network property, we divide the devices into different clusters according to how many times they can traverse the whole Markov process. For example, one device will be allocated into the cluster $\mathbb{C}_i$ if it can traverse the Markov process for $l_i$ times. Here $l_i$ can be expressed as
\begin{equation}
l_i = \left\lfloor \frac{{T_i}}{\Delta} \right\rfloor = \left\lfloor \frac{{2Rcos\theta_i}}{v*\Delta} \right\rfloor,
\end{equation}
where $\left\lfloor x \right\rfloor$ represents the rounding down to $x$. $\Delta$ is the time for a packet to traverse all the Markov stages. According to \cite{xiao}, $\Delta$ is shown as
\begin{equation}\label{deta}
\begin{split}
\Delta  = &E({B_i})\delta  + E({F_i})[\frac{{{P_{si}}}}{{{P_{bi}}}}{T_s} + \frac{{({P_{bi}} - {P_{si}})}}{{{P_{bi}}}}{T_c}] \\&+ {L}({T_c} + {T_o}),
\end{split}
\end{equation}
where ${P_{si}}$ and ${P_{bi}}$ are the probabilities that the devices in $\mathbb{C}_i$ transmit a packet successfully and the probability that there is at least one device transmitting the packet, respectively. ${B_i}$ represents the total number of backoff counter and $E({B_i})$ is the average number of backoff counter during the backoff stage, which can be shown as
\begin{equation}
E({B_i}) = \sum\limits_{j = 0}^{{L}} {\frac{{{W_{i,j}} - 1}}{2}}.
\end{equation}

Assume that ${F_i}$ is the overall time when the counter freezes and $E({F_i})$ is shown as
\begin{equation}
E({F_i}) = \frac{{E({B_i})}}{{1 - {q_i}}}{q_i}.
\end{equation}

Let $T_{s}$ and $T_{c}$ in \eqref{deta} denote the average time that the channel is sensed busy and the average time for a collision. $T_{o}$ is the time that a device has to wait after the access collision before the next channel sensing.

Here the basic access mechanism is a two-way handshaking technique while the RTS/CTS is a four-way
handshaking technique, which are the most common mechanisms in 802.11 \cite{b4} and \cite{b11}. 
The definitions of $T_s$, $T_c$ and $T_o$ are different in the basic access mechanism and RTS/CTS mechanism. For clarity, we use $T_s^{b}$, $T_c^{b}$, $T_o^{b}$ and $T_s^{r}$, $T_c^{r}$, $T_o^{r}$ to denote the parameters in basic access mechanism and RTS/CTS mechanism, respectively. Then, we have
\begin{equation}
T_s^{b}=T_{\rm{H}}+T_{\rm{E}}+\rm{SIFS}+T_{\rm{ACK}}+\rm{DIFS}+2*\delta,
\end{equation}
\begin{equation}
T_c^{b}=T_{\rm{H}}+T_{\rm{E}}+\rm{DIFS}+\delta,
\end{equation}
\begin{equation}
T_o^{b}=\rm{SIFS}+T_{\rm{ACK\_timeout}},
\end{equation}
\begin{equation}
\begin{split}
T_s^{r}=&T_{\rm{RTS}}+T_{\rm{CTS}}+T_{\rm{H}}+T_{\rm{E}}+3*\rm{SIFS}+\\& T_{\rm{ACK}}+\rm{DIFS},
\end{split}
\end{equation}
\begin{equation}
T_c^{r}=T_{\rm{RTS}}+\rm{SIFS}+T_{\rm{ACK}}+\rm{DIFS},
\end{equation}
\begin{equation}
T_o^{r}=\rm{SIFS}+T_{\rm{CTS\_timeout}},
\end{equation}
where $T_{\rm{H}}$, $T_{\rm{E}}$, $T_{\rm{ACK}}$, SIFS, DIFS, $T_{\rm{RTS}}$ and $T_{\rm{CTS}}$ denote the time to transmit the header (including MAC header, physical header), the time to transmit a payload with length E, the time to transmit an ACK, the time durations of a SIFS and DIFS, and the time to transmit a RTS and CTS , respectively. $T_{\rm{CTS\_timeout}}$ and $T_{\rm{ACK\_timeout}}$ denote the duration of the ACK and CTS timeouts respectively. $\delta$ denotes the duration for an idle time slot.

Under this division strategy, the devices located in the center band of the UBS's coverage can connect with the UBS for the longest time. If these devices repeat the backoff process for $N$ times, they belong to the cluster $\mathbb{C}_N$ and $N=\left\lfloor \frac{t_N}{v*\Delta}\right\rfloor$.
In this way, all the devices in the coverage can be divided into $N$ clusters and each devices in the cluster $\mathbb{C}_i, (i=1,2,...,N)$ can repeat the Markov process for $i$ times.

On the other hand, we assume the devices are uniformly distributed in the UBS's coverage with density $\rho$, the number of the devices in the area follows the Poission distribution according to \cite{b16}. The probability that there are $n$ devices in the cluster $\mathbb{C}_i$ can be shown as
\begin{equation}\label{eq8}
{f_i}(n) = \frac{{{{(\rho {A_i})}^n}{e^{ - \rho {A_i}}}}}{{n!}},n = 0,1,2,...,\infty,
\end{equation}
where ${A_i}$ is the area of $\mathbb{C}_i$. This distribution can help us to analyze the saturation throughput of the network in the following section.

\section{Performance Analysis}\label{section:3}


This section analyzes the Markov chain process and gives the solution of the saturation throughput.

\subsection{Markov Chain Process Analysis} \label{Markov}
Before the analysis, we first define the stationary distribution of the chain ${b_{i,j,k}}$ as
\begin{equation}
\begin{split}
{b_{i,j,k}} = & \mathop {\lim }\limits_{t \to \infty } P\{ cluster=i,s(t) = j,b(t) = k\},\\& ~j \in [0,L],~k \in [0,{W_{i,j}}].
\end{split}
\end{equation}

Then, for a device in the cluster $i$, we get
\begin{equation}\label{eq6}
{b_{i,j,0}} = {P_{eq}}{b_{i,j - 1,0}} = P_{eq}^j{b_{i,0,0}}, ~{\rm{  j}} \in {\rm{[0,}}{{\rm{L}}_{\rm{i}}}{\rm{]}},
\end{equation}
where $P_{eq}$ can be expressed as
 \begin{equation}\label{eq6-1}
P_{eq} = (1 - {Q_i}){q_i} + {Q_i},
\end{equation}
where $q_i$ is the probability that the channel is busy when the device is in the backoff stage. During the access process, according to the CSMA/CA protocol, once a device in the cluster $\mathbb{C}_i$ connects to the UBS, other devices either in the cluster $\mathbb{C}_i$ or not cannot transmit information to the UBS. Combining the Poission distribution of the number of the devices in \eqref{eq8}, we can calculate the $q_i$ as \eqref{long_1}.


According to the chain regularities, each value of the backoff counter $k \in [1,{W_{i,j}} - 1]$ can be obtained by \eqref{long_2}.

By substituting \eqref{eq6} in \eqref{long_2}, all the values ${b_{i,j,k}}$ can be expressed as a function of ${b_{i,0,0}}$. Then, considering the normalization condition, we can easily obtain:
\setcounter{equation}{20}
\begin{equation}\label{eq12}
\begin{split}
1&= \sum\limits_{j = 0}^{{L}}\sum\limits_{k = 0}^{{W_{i,j}} - 1} {{b_{i,j,k}}} \\&= \sum\limits_{j = 0}^{{L}} {{b_{i,j,0}}}  + \sum\limits_{k = 1}^{{W_{i,j}} - 1} {{b_{i,0,k}}}  + \sum\limits_{j = 1}^{{L}} \sum\limits_{k = 1}^{{W_{i,j}} - 1} {{b_{i,j,k}}},
\end{split}
\end{equation}
where ${P_a} = (1 - {P_{eq}}), {P_b} = (1 - 2{P_{eq}})$. Therefore, we can obtain the expression of $b_{i,0,0}$, as shown in \eqref{long_3}.

According to the Markov chain, the devices stop the transmission when their backoff counters decrease to 0. Thus, the probability $\tau_i$ that one device in the cluster $i$ transmits a packet in a randomly selected time slot can be shown as:
\setcounter{equation}{22}
\begin{equation}\label{eq111}
{\tau _i} = \sum\limits_{j = 0}^{{L}} {{b_{i,j,0}} = \frac{{1 - P_{eq}^{{L} + 1}}}{{1 - {P_{eq}}}}{b_{i,0,0}}}.
\end{equation}

By combining \eqref{long_1} and \eqref{eq111}, we can calculate the value of parameters $\tau_i$ numerically.

\subsection{Saturation Throughput Computation}

Let $P_{tr}$ be the probability that there is at least one device doing the transmission for a certain time. According to \cite{b6}, it can be shown as
\begin{equation}\label{eq17}
{P_{tr}} = 1 - \prod\limits_{h = 1}^N {(\sum\limits_{n = 0}^\infty  {{f_h}(n)} {{(1 - {\tau _h})}^n})}.
\end{equation}

For the devices in cluster $\mathbb{C}_i$, the probability of successful transmission $P_{s_i}$ can be calculated as
\begin{equation}\label{eq18}
\begin{split}
{P_{{s_i}}} =& \left[\sum\limits_{n = 1}^\infty  {{f_h}(n)} n{\tau _i}{(1 - {\tau _i})^{n - 1}}\right] \times \\& \prod\limits_{h = 0,h \ne i}^{N - 1} \left[{\sum\limits_{n = 0}^\infty  {{f_h}(n)} {{(1 - {\tau _h})}^n}} \right].
\end{split}
\end{equation}

Therefore, in the whole coverage, the overall probability ${P_s}$ of successful packet transmission to UBS can be shown as ${P_s} = \sum\limits_{i = 1}^{N} {{P_{{s_i}}}}$. Finally, by combining \eqref{eq17} and \eqref{eq18}, the throughput of the network can be shown as
\begin{equation}\label{eq19}
S = \frac{{{P_s}{P_{tr}}E[P]}}{{(1 - {P_{tr}})
		\sigma  + {P_s}{P_{tr}}{T_s} + {P_{tr}}(1 - {P_s}){T_c}}}.
\end{equation}

\section{Simulation and Numerical Results} \label{section:5}

In this section, we validate our system model firstly by comparing the theoretical results with those obtained by the Monte Carlo simulation. Then, the effects of the network parameters, such as UBS's velocity and the density of the devices, on the throughput are analyzed. An idle time slot $\delta $ is 50$\mu s$, ${\rm{ACK}_{\rm{timeout}}}$= 300 $\mu s$, ${\rm{CTS}_{\rm{timeout}}}$= 300 $\mu s$, ACK= 112 $\mu s$, RTS= 160 $\mu s$, CTS= 112 $\mu s$, SIFS=28 $\mu s$, DIFS= 128 $\mu s$. Assume the channel data rate is equal to $1 Mbit/s$ and the frame size , denoted by EP, is fixed at $8k$ Bytes. The radius of the coverage is assumed to be 1000m unless stated otherwise.

\begin{figure}[t]
	\centering
	\includegraphics [width=8cm]{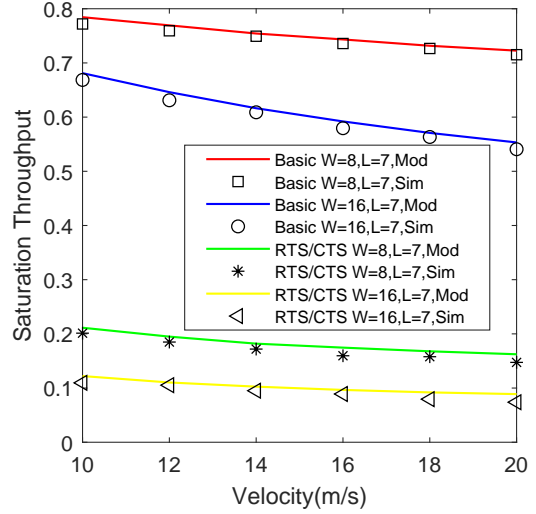}
	\caption{Saturation throughput vs. velocity of the UBS.}
	\label{validation}
\end{figure}

Fig.~\ref{validation} indicates the normalized saturation throughput versus the velocity of the UBS under different parameters for the theoretical model and the simulation, marked as `Mod' and `Sim' in legends respectively.
To fully verify the network performance, we validate the results in both basic access and RTS/CTS cases. The user density in this simulation is $\rho=50/{km}^2$ and the retry limit is $L=7$ for all devices. In addition, we adopt two values of the initial contention window size ($CW_{min}=8,16$) to explore the network performance. It is clear to see that the theoretical results (lines) match the simulation results (symbols) well, which perfectly confirm our model. Moreover, the saturation throughput of proposed model decreases with the increasing of the UBS's velocity. This is because the device can connect with the UBS for a shorter time when the UBS flies faster. It is hard for all the devices to access the channel and fewer packets can be transmitted successfully because of the short communication duration, leading to the the lower throughput.

\begin{figure}[htbp]
	\centering\subfigure[Density of the devices.]{\includegraphics[width=4.55cm]{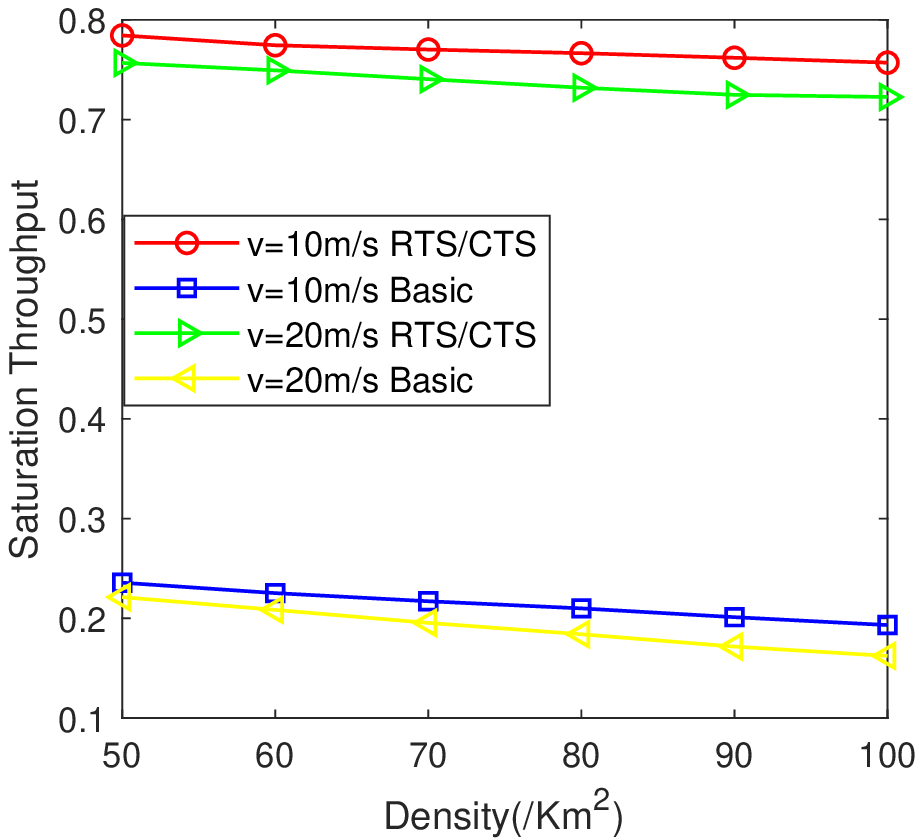}\label{den_basic}}
	\hspace{-5mm}
	\centering\subfigure[Retry limit.]{\includegraphics[width=4.55cm]{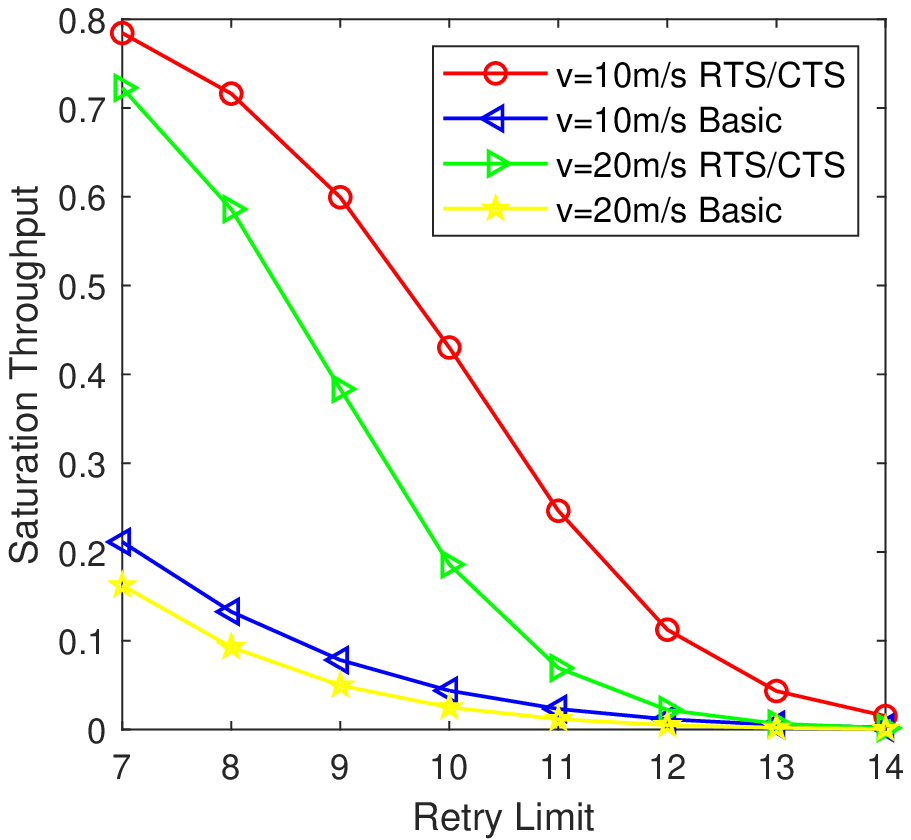}\label{re_basic}}
	\caption{Saturation throughput vs. device density and retry limit.}
\end{figure}

In Fig.~\ref{den_basic}, we plot the normalized saturation throughput vs. the density of the devices for both basic access and RTS/CTS mechanisms under two different UBS's velocities ($v=10, 20m/s$).
The density of the devices increases from $50/{km}^2$ to $100/{km}^2$.
It can be seen that the density of the devices influences the performance negatively. The throughput decreases along with the increase of the devices' density.
 This is because with the constraint of the spectrum resource, the more devices participate in the access process at the same time, the more access collisions will happen, which makes it harder to transmit a packet successfully and result in the lower throughput.
 In addition, by comparing the throughput of those two mechanisms, we can find that the RTS/CTS performs better than the basic access mechanism in saturation throughput.
 Compared with the basis mechanism, RTS/CTS is designed to combat the problem of hidden terminal and it can reduce the duration of a collision to increase the system performance.
 Therefore, RTS/CTS is more likely to give better performance than the basic one especially when long messages are transmitted.


In the simulations, we also implement some experiments to analyze the influence of the retry limit on the saturation throughput for both basic access and RTS/CTS mechanisms, shown in Fig.~\ref{re_basic}.
 From Fig.~\ref{re_basic}, we can see that the increase of the retry limit (from 7-14) has negative impact on the throughput. It is because that the devices have more opportunities to retry the transmission with larger retry limit, which will cause more collisions and it is more difficult to transmit a packet successfully in the network. As a result, the throughput will be influenced negatively.

\begin{figure}[htbp]
	\centering\subfigure[Initial contention window size.]{\includegraphics[width=4.55cm]{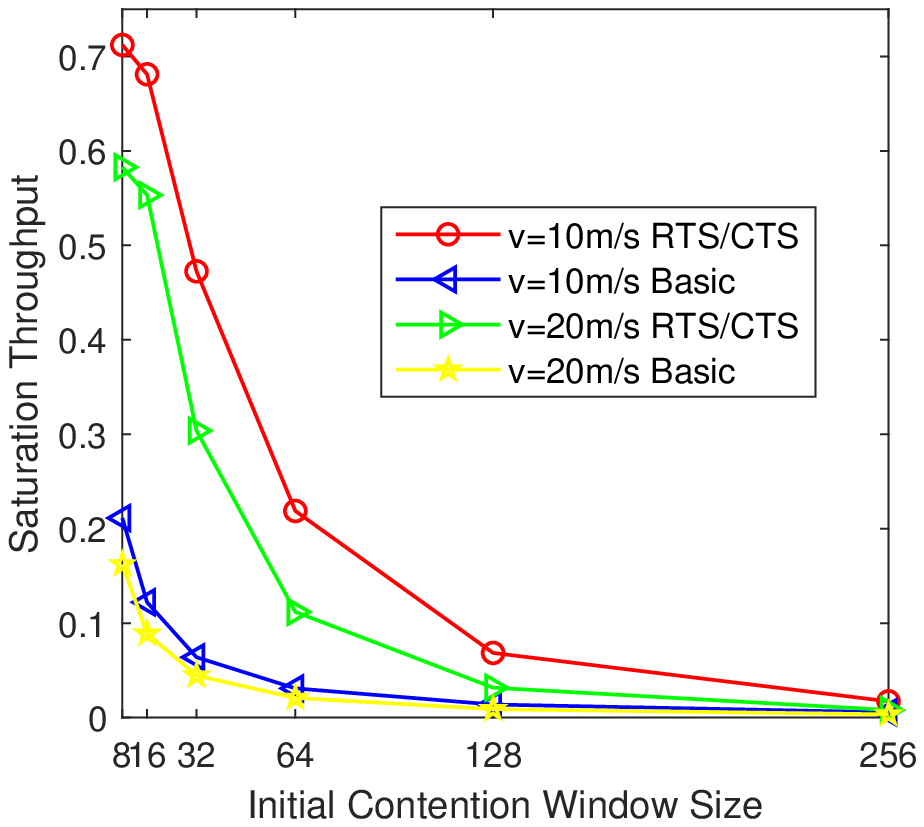}\label{cw}}
	\hspace{-5mm}
	\centering\subfigure[UBS's coverage.]{\includegraphics[width=4.55cm]{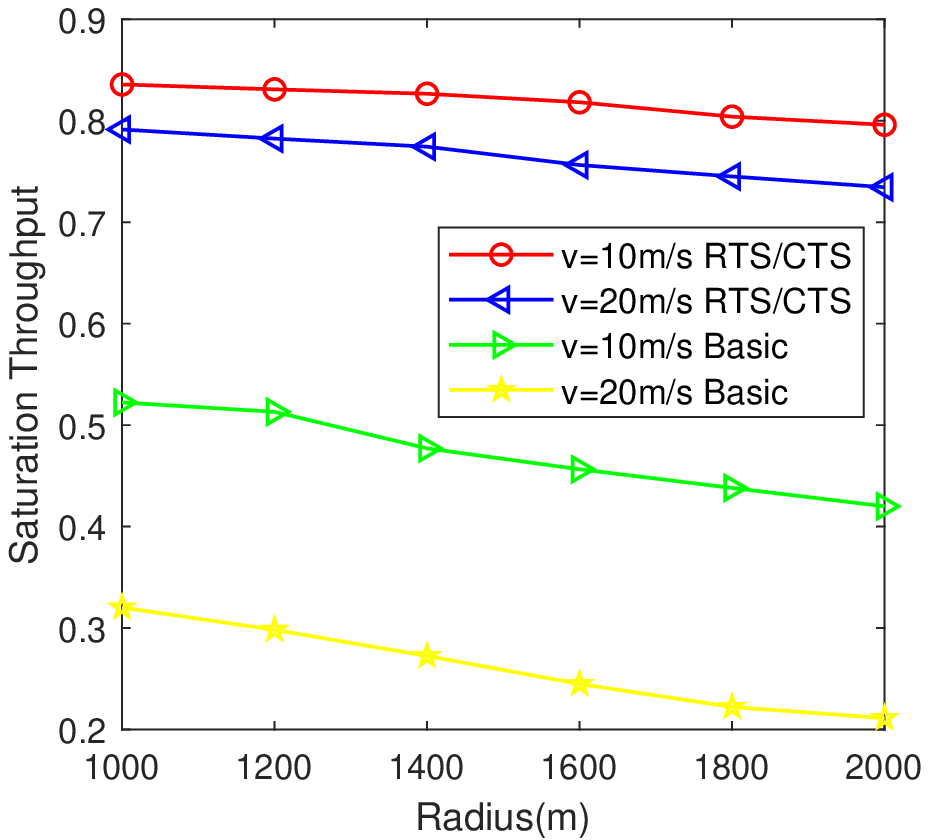}\label{r_basic}}
	\caption{Saturation throughput vs. initial contention window size and UBS's coverage.}
\end{figure}

The impacts of the initial contention window size are also analyzed in Fig.~\ref{cw}.
According to Fig.~\ref{cw}, we can conclude that the effects of the initial contention window size is serious. For basic access mechanism, the throughput declines to almost zero when the initial contention window size climbs to 256. For RTS/CTS mechanism, the effects cannot be ignored either. The throughput drops to below 0.1 in this scenario. The reason is that if the initial contention window increases, the devices will spend more time in the backoff process. Therefore, the channel is more likely to stay in idle state, causing the waste for the part of the channel resource. On the other hand, the transmission delay will be longer for larger window size which also influences the throughput performance.


Finally, the impacts of UBS's coverage are considered. Since the UBS's coverage is a circle, we just analyzes the effects of the radius of the coverage. Here we set the density of the devices is $\rho=50/{km}^2$ and the radius of the coverage increases from 1000 m to 2000 m. The results for basic access mechanism and RTS/CTS mechanism are given in Fig~\ref{r_basic}.

The throughput drops with the increasing of the radius both in basic access mechanism and RTS/CTS mechanism. For larger coverage area, more devices will connect with the UBS and transmit the packet simultaneously, leading to the more congestion. Accordingly, the system's throughput decreases both for basic and RTS/CTS mechanisms. Though both increasing the density of the devices and UBS's coverage radius enlarge the number of the devices that the UBS serves, there is sitll some difference between these two methods. The larger coverage not only increases the number of the devices, but also improves the communication duration between the devices and the UBS. Thus, the results are different between Fig.~\ref{den_basic} and Fig.~\ref{r_basic}.

\section{Conclusions}\label{section:6}
In our paper, a scenario where a UBS flies straightly to collect data from a set of IoT devices is considered. Based on the characteristics of the UAV-based network, we propose a quitting probability and develop a Markov chain model to calculate the saturation throughput of CSMA/CA in this UBS data collection system. In addition, we also analyze the impacts of different network parameters including retry limit, initial contention window size, UBS's velocity, the density of the devices and the UBS's coverage.

\bibliographystyle{ieeetran}
\bibliography{tvt}

\begin{thebibliography}{10}
\providecommand{\url}[1]{#1}
\csname url@samestyle\endcsname
\providecommand{\newblock}{\relax}
\providecommand{\bibinfo}[2]{#2}
\providecommand{\BIBentrySTDinterwordspacing}{\spaceskip=0pt\relax}
\providecommand{\BIBentryALTinterwordstretchfactor}{4}
\providecommand{\BIBentryALTinterwordspacing}{\spaceskip=\fontdimen2\font plus
\BIBentryALTinterwordstretchfactor\fontdimen3\font minus
  \fontdimen4\font\relax}
\providecommand{\BIBforeignlanguage}[2]{{%
\expandafter\ifx\csname l@#1\endcsname\relax
\typeout{** WARNING: IEEEtran.bst: No hyphenation pattern has been}%
\typeout{** loaded for the language `#1'. Using the pattern for}%
\typeout{** the default language instead.}%
\else
\language=\csname l@#1\endcsname
\fi
#2}}
\providecommand{\BIBdecl}{\relax}
\BIBdecl

\bibitem{b1}
L.~{Gupta}, R.~{Jain}, and G.~{Vaszkun}, ``Survey of important issues in uav
  communication networks,'' \emph{IEEE Communications Surveys Tutorials},
  vol.~18, no.~2, pp. 1123--1152, Secondquarter 2016.

\bibitem{b2}
Y.~{Zeng}, R.~{Zhang}, and T.~J. {Lim}, ``Wireless communications with unmanned
  aerial vehicles: opportunities and challenges,'' \emph{IEEE Communications
  Magazine}, vol.~54, no.~5, pp. 36--42, May 2016.

\bibitem{d1}
D.~{Yang}, Q.~{Wu}, Y.~{Zeng}, and R.~{Zhang}, ``Energy tradeoff in
  ground-to-uav communication via trajectory design,'' \emph{IEEE Transactions
  on Vehicular Technology}, vol.~67, no.~7, pp. 6721--6726, July 2018.

\bibitem{d2}
Q.~{Wu}, Y.~{Zeng}, and R.~{Zhang}, ``Joint trajectory and communication design
  for multi-uav enabled wireless networks,'' \emph{IEEE Transactions on
  Wireless Communications}, vol.~17, no.~3, pp. 2109--2121, March 2018.

\bibitem{d3}
B.~Li, Y.~Jiang, J.~Sun, L.~Cai, and C.-Y. Wen, ``Development and testing of a
  two-uav communication relay system,'' \emph{Sensors}, vol.~16, no.~10, p.
  1696, 2016.

\bibitem{d4}
Y.~{Li} and L.~{Cai}, ``Uav-assisted dynamic coverage in a heterogeneous
  cellular system,'' \emph{IEEE Network}, vol.~31, no.~4, pp. 56--61, July
  2017.

\bibitem{ba}
M.~{Mozaffari}, W.~{Saad}, M.~{Bennis}, and M.~{Debbah}, ``Drone small cells in
  the clouds: Design, deployment and performance analysis,'' in \emph{2015 IEEE
  Global Communications Conference (GLOBECOM)}, Dec 2015, pp. 1--6.

\bibitem{b3}
M.~{Mozaffari}, W.~{Saad}, and M.~{Bennis}, ``Mobile unmanned aerial vehicles
  (uavs) for energy-efficient internet of things communications,'' \emph{IEEE
  Transactions on Wireless Communications}, vol.~16, no.~11, pp. 7574--7589,
  Nov 2017.

\bibitem{b6}
G.~{Bianchi}, ``Performance analysis of the ieee 802.11 distributed
  coordination function,'' \emph{IEEE Journal on Selected Areas in
  Communications}, vol.~18, no.~3, pp. 535--547, March 2000.

\bibitem{b7}
F.~{Daneshgaran}, M.~{Laddomada}, F.~{Mesiti}, and M.~{Mondin}, ``Unsaturated
  throughput analysis of ieee 802.11 in presence of non ideal transmission
  channel and capture effects,'' \emph{IEEE Transactions on Wireless
  Communications}, vol.~7, no.~4, pp. 1276--1286, April 2008.

\bibitem{xiao}
{Yang Xiao}, ``Performance analysis of priority schemes for ieee 802.11 and
  ieee 802.11e wireless lans,'' \emph{IEEE Transactions on Wireless
  Communications}, vol.~4, no.~4, pp. 1506--1515, July 2005.

\bibitem{b4}
D.~{Ho}, J.~{Park}, and S.~{Shimamoto}, ``Performance evaluation of the pfsc
  based mac protocol for wsn employing uav in rician fading,'' in \emph{2011
  IEEE Wireless Communications and Networking Conference}, March 2011, pp.
  55--60.

\bibitem{b11}
P.~{Chatzimisios}, A.~C. {Boucouvalas}, and V.~{Vitsas}, ``Performance analysis
  of ieee 802.11 dcf in presence of transmission errors,'' in \emph{2004 IEEE
  International Conference on Communications (IEEE Cat. No.04CH37577)}, vol.~7,
  June 2004, pp. 3854--3858 Vol.7.

\bibitem{b16}
B.~{Li}, H.~{Li}, W.~{Wang}, Z.~{Hu}, and Q.~{Yin}, ``Energy-effective relay
  selection by utilizing spacial diversity for random wireless sensor
  networks,'' \emph{IEEE Communications Letters}, vol.~17, no.~10, pp.
  1972--1975, October 2013.

\end{thebibliography}

\end{document}